\documentclass[aps,preprint,superscriptaddress,superscriptemail,showpacs,showkeys]{revtex4}
\usepackage[dvipdf]{graphicx}
\usepackage{amssymb}
\usepackage{amsbsy}
\usepackage{latexsym}
\usepackage{epsfig}

\newcommand{\be}{\begin{equation}}
\newcommand{\ee}{\end{equation}}
\newcommand{\ba}{\begin{eqnarray}}
\newcommand{\ea}{\end{eqnarray}}

\newcommand{\dg}{\dagger}
\newcommand{\wbe}{\begin{widetext}}
\newcommand{\wee}{\end{widetext}}

\begin{document}

\title{BRST quantization of a sixth-order derivative scalar field theory}

\author{Yong-Wan Kim}

\email{ywkim65@gmail.com}

\affiliation {Center for Quantum Spacetime, Sogang University,
Seoul 121-742, Korea}

\author{Yun Soo Myung}

\email{ysmyung@inje.ac.kr}

\affiliation {Institute of Basic Science and School of Computer
Aided Science, Inje University, Gimhae 621-749, Korea}

\author{Young-Jai Park}

\email{yjpark@sogang.ac.kr}

\affiliation {Center for Quantum Spacetime, Sogang University,
Seoul 121-742, Korea}

\affiliation {Department of Physics, Sogang University, Seoul
121-742, Korea}

\begin{abstract}
We study a sixth order derivative scalar field model in Minkowski
spacetime as a toy model of higher-derivative critical gravity
theories. This model is consistently quantized when using the
Becchi-Rouet-Stora-Tyutin (BRST) quantization scheme even though it
does not show gauge symmetry manifestly. Imposing a BRST quartet
generated by two scalars and ghosts, there remains a non-trivial
subspace with positive norm. This might be interpreted as a
Minkowskian dual version of the unitary truncation in the
logarithmic conformal field theory.

\end{abstract}



\maketitle

\section{Introduction}
Stelle~\cite{Stelle} has first introduced the quadratic curvature
gravity  of $\alpha(R_{\mu\nu}^2-R^2/3)+\beta R^2$  to improve  the
perturbative properties of Einstein gravity. In case of
$\alpha\beta\not=0$, the renormalizability was achieved but the
unitarity was violated for $\alpha\not=0$, showing that the
renormalizability and unitarity exclude to each other. Although the
$\alpha$-term of providing  massive graviton improves the
ultraviolet divergence, it induces ghost excitations which spoil the
unitarity.  The price one has to pay for making the
theory renormalizable is the loss of unitarity.

After this work, a first requirement  for the quantum gravity is to
gain the unitarity, which means that its linearized theory has no
tachyon and ghost in the particle content~\cite{Barth:1983hb}. To
that end, critical gravities have recently  received  much attention
because they were considered as toy models for quantum
gravity~\cite{Li:2008dq,Lu:2011zk,Deser:2011xc,Porrati:2011ku,
Alishahiha:2011yb,Bergshoeff:2011ri,Myung:2011bn,Lu:2011mw}.  At the
critical point, a degeneracy took  place in the AdS spacetime and
thus, ghost-like massive gravitons become massless gravitons.
Instead of massive gravitons, an equal amount of logarithmic modes
were introduced for the critical gravity. However, one has to
resolve the non-unitarity problem of the critical gravity theories
because these contain higher-derivative
interactions~\cite{Lu:2011zk}.  It was shown that a rank-2
logarithmic conformal field theory (LCFT) is dual to a critical
gravity~\cite{Grumiller:2008qz,Myung:2008dm,Maloney:2009ck}. Thus,
the non-unitarity of critical gravity is closely related to that  of
the rank-2 LCFT  where the Hamiltonian cannot be diagonalized on the
fields due to the Jordan
structure~\cite{Gurarie:1993xq,Flohr:2001zs}.

In order to address the non-unitarity issue of critical gravity, it
has been proposed to truncate log-modes out by imposing the AdS
boundary conditions~\cite{Bergshoeff:2012sc}. A rank of the LCFT
refers to the dimensionality of the Jordan cell. The rank-2 LCFT
dual to a critical gravity has a rank-2 Jordan cell and thus, an
operator has a logarithmic partner. However, there remains nothing
for the rank-2 LCFT after truncation. Dipole-ghost fields ($A,B$) on
AdS$_3$ space are also dual to the rank-2
LCFT~\cite{Kogan:1999bn,Myung:1999nd}, whereas they form a BRST
quartet to give zero norm
state~\cite{Kugo:1979gm,Becchi:1975nq,Tyutin:1975qk} after
introducing ghosts in Minkowski spacetime~\cite{Rivelles:2003jd}.
Instead, a polycritical gravity was introduced to provide a
polycritical point~\cite{Nutma:2012ss,Kleinschmidt:2012rs} whose
dual is supposed to be  a higher rank LCFT.     The LCFT dual to a
tricritical gravity has rank-3 Jordan cell~\cite{Bergshoeff:2012ev}
and  an operator has two logarithmic partners of log and log$^2$.
After truncation, there remains a unitary subspace with non-negative
inner product. Its dual scalar model was investigated on the BTZ
black hole spacetime explicitly~\cite{Moon:2012vc}.

In this direction, it is very important to understand the truncation
mechanism which  leads naturally to  the unitary subspace. However,
the fact that the bulk spacetime is AdS and theory is a polycritical
gravity prevents us from understanding this truncation scheme well.
Hence, we consider  a sixth-order derivative scalar field theory in
Minkowski spacetime. To avoid a difficulty in dealing with a single
sixth-order derivative theory directly~\cite{de Urries:1998bi}, we
introduce an equivalent three coupled scalar fields with degenerate
masses.
 This model will be  quantized by  employing  the
BRST quantization scheme even though it does not show gauge symmetry
manifestly.  Imposing a BRST quartet generated by two scalars
$\phi_1,\phi_3$ and ghosts $c,d$, there remains a non-trivial
subspace with positive norm for $\phi_2$. This could be interpreted
to  be a Minkowskian dual version of the unitary truncation in the
LCFT.

Our  action consists  of three scalar fields $\phi_1$, $\phi_2$,
$\phi_3$, and ghost fields $c$, $d$ with degenerate  masses
$(m_1=m_2=m$) in four dimensional spacetime
 \ba \label{action1}
 S=-\int d^4x\left[ \partial_\mu\phi_1\partial^\mu\phi_3+\frac{1}{2}\partial_\mu\phi_2\partial^\mu\phi_2
                  +\phi_1\phi_2
                  +m_1^2\phi_1\phi_3+\frac{1}{2}m_2^2\phi^2_2
                  +\partial_\mu c \partial^\mu d +m_1^2 cd     \right].
 \ea
 Without ghost term, this action appeared
 in~\cite{Bergshoeff:2012sc,Moon:2012vc}.
Here, we adopt the conventions of $\eta_{\mu\nu}={\rm diag.}(-+++)$
and  $x^\mu=(t,\vec{x})$. We do not consider the case of
non-degenerate masses ($m_1\not=m_2$) because we may consider
(\ref{action1}) without ghosts as a toy model of tricritical gravity
in Minkowski spacetime. We note that the action (\ref{action1}) is
invariant under BRST transformations
 \be \label{brstra}
 \delta\phi_1=0,~~ \delta\phi_2=0,~~ \delta\phi_3=c,~~ \delta
 c=0,~~ \delta d=\phi_1.
 \ee
 Here the ghost numbers are assigned to be $[\phi_i]=0~(i=1,2,3),~[c]=-1,$ and $[d]=1$.
For the case of  non-degenerate masses, there is no  BRST invariance
like (\ref{brstra}) because this symmetry is not nilpotent.

\section{Sixth-order time derivative particle theory}
In order to understand a sixth-order derivative scalar theory,  we
consider first its sixth-order time derivative version of the action
(\ref{action1})
 \ba\label{action2}
 S=\int dt
 \left[\dot{\phi}_1\dot{\phi}_3+\frac{1}{2}\dot{\phi}^2_2-\phi_1\phi_2-m^2\phi_1\phi_3-\frac{1}{2}m^2\phi^2_2
      +\dot{c}\dot{d}-m^2cd  \right],
 \ea
where the dot denotes  differentiation with respect to time. This
action could  describe a  sixth order  derivative harmonic
oscillator for $\phi_3$ when coupling to $\phi_1$ and $\phi_2$.
Equations of motion can be obtained as
 \ba
 \label{eomp1}\ddot{\phi}_1+m^2\phi_1&=&0,\\
 \label{eomp2}\ddot{\phi}_2+m^2\phi_2&=&-\phi_1,\\
 \label{eomp3}\ddot{\phi}_3+m^2\phi_3&=&-\phi_2
 \ea
by varying $\phi_3$, $\phi_2$, and $\phi_1$,
respectively. Eliminating  $\phi_1$ and $\phi_2$ leads to  the
the sixth-order time derivative equation for $\phi_3$ as
 \be\label{6theq}
 \left(\frac{d^2}{dt^2}+m^2\right)^3\phi_3=0.
 \ee
On the other hand, eliminating $\phi_1$ leads to the fourth order
equation for $\phi_2$ \be\label{4theq}
 \left(\frac{d^2}{dt^2}+m^2\right)^2\phi_2=0
 \ee
which is recognized to be the degenerate  Pais-Uhlenbeck oscillator
with $m_i=\omega_i$ \cite{Pais:1950za}. Its field theory realization
was reported in~\cite{Jimenez:2012ak}.

\begin{figure}[t!]
   \centering
   \epsfbox{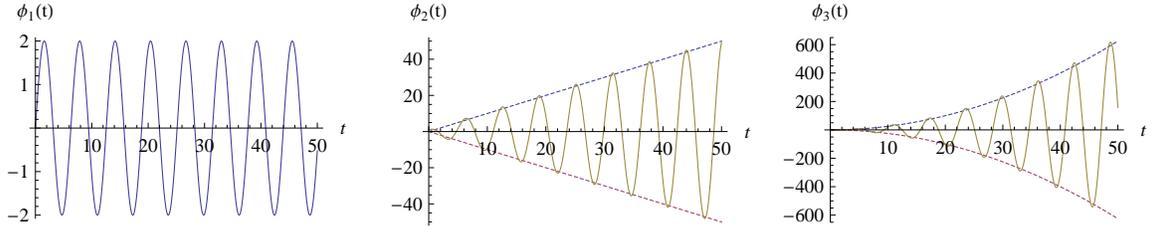}
\caption{Temporal behaviors of the solutions of $\phi_1$, $\phi_2$,
and $\phi_3$ with $N_1$=$m$=$a_i$=$a^\dg_i$=1 for $i=1,~2,~3$.}
\label{fig1}
\end{figure}

The solutions to Eqs.~(\ref{eomp1})-(\ref{eomp3}) are given by
 \ba
 \label{sol1}\phi_1(t)&=&iN_1\left(a_1e^{-imt}-a^\dg_1e^{imt}\right),\\
 \label{sol2}\phi_2(t)&=&\frac{N_1}{2m^2}\left[\left(a_2+a_1mt\right)e^{-imt}+\left(a^\dg_2+a^\dg_1mt\right)e^{imt}\right],\\
 \label{sol3}\phi_3(t)&=&-\frac{iN_1}{4m^4}\left[\left(a_3+\left(a_2-\frac{i}{2}a_1\right)mt+\frac{1}{2}a_1m^2t^2\right)e^{-imt}\right.\nonumber\\
        &&~~~~~~~~\left.-\left(a^\dg_3+\left(a^\dg_2+\frac{i}{2}a^\dg_1\right)mt+\frac{1}{2}a^\dg_1m^2t^2\right)e^{imt}\right],
 \ea
respectively. Note that they are all hermitians. It is also easy to
check  that the solution (\ref{sol3})[(\ref{sol2})] solves the
higher-order equation (\ref{6theq})[(\ref{4theq})] directly.
Fig.~\ref{fig1} shows the temporal behaviors of the solutions.
$\phi_1$ shows  a pure oscillation in time.  Even though $\phi_2$
and $\phi_3$ are growing linearly and quadratically in time, their
growths are milder than an exponentially growing mode. Although
these linear and quadratic growths are independent of the presence
of the Ostrograski instability,  they  show that $\phi_3$ and
$\phi_2$ are solutions to higher-order time derivative equations
(\ref{6theq}) and (\ref{4theq}).

\begin{figure}[t!]
   \centering
   \epsfbox{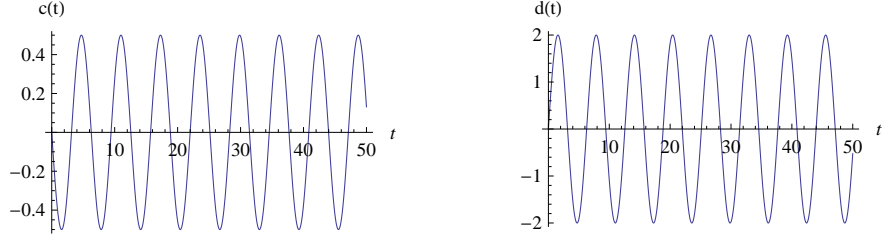}
\caption{Temporal behaviors of the ghosts of  $c(t)$ and  $d(t)$
with $N_1$=$m$=$c_1$=$c^\dg_1$=$d_1$=$d^\dg_1$=1.} \label{fig2}
\end{figure}

On the other hand, equations of motion for the ghosts
 \ba
 && \ddot{c}+m^2c=0,\\
 && \ddot{d}+m^2d=0
 \ea
lead to solutions as
 \ba
 c(t)&=&-\frac{iN}{4m^4}\left(c_1e^{-imt}-c^\dg_1e^{imt}\right),\\
 d(t)&=&iN_1\left(d_1e^{-imt}-d^\dg_1e^{imt}\right),
 \ea
respectively. Fig.~\ref{fig2} shows the temporal behaviors of the
ghost solutions as $\phi_1$ does show.

Now we are in a position to carry out canonical quantization of
(\ref{action2}). This may provide a hint to quantization of its full
action (\ref{action1}). Canonical quantization can be started with  finding canonical momenta
 \be\label{canmom}
 \pi_1=\dot{\phi}_3,~~~\pi_2=\dot{\phi}_2,~~~\pi_3=\dot{\phi}_1.
 \ee
Canonical Hamiltonian is given by
 \be \label{chamil}
 {\cal H}_c=\pi_1\pi_3+\frac{1}{2}\pi^2_2+\phi_1\phi_2+m^2\phi_1\phi_3+\frac{1}{2}m^2\phi^2_2.
 \ee
Expressing (\ref{chamil}) in terms of modes $a_1$, $a_2$, and $a_3$
in Eqs.~(\ref{sol1})-(\ref{sol3}), the canonical Hamiltonian becomes
 \be\label{canH_ptls}
 {\cal H}_c=\frac{N^2_1}{2m^2}\left[a^\dg_2a_2+2i\left(a^\dg_2a_1-a^\dg_1a_2\right)-a^\dg_3a_1-a^\dg_1a_3\right].
 \ee
From now on, we would take $N^2_1=2m^3$ for convenience, if not
mentioned otherwise.  The equal-time commutation relations between
operators are obtained as
 \ba
 [a_1,a^\dg_3]=-1,~~[a_2,a^\dg_2]=1,~~[a_2,a^\dg_3]=i,~~[a_3,a^\dg_3]=\frac{3}{2}.
 \ea
These can be cast into the following matric form:
 \be
 [a_i, a^\dg_j]=
 \left(
  \begin{array}{ccc}
   0 & 0 & -1 \\
    0 & 1 & i \\
    -1 & -i & \frac{3}{2} \\
  \end{array}
 \right).
 \ee
Although its Hamiltonian is not diagonal and positive definite,
their commutation relations reveal an useful information between
operators. Clearly, $[a_2,a^\dg_2]=1$ represents a standard
commutation relation, while others do show non-standard
commutations. In order to make the Hamiltonian diagonal, one has to
introduce new operators $b_i$ by using  transformations
 \ba
 a_1&=&\frac{3\sqrt{2}i}{8}b_1-\frac{\sqrt{2}i}{8}b_2+\frac{i}{2}b_3,\nonumber\\
 a_2&=&\sqrt{2}b_1-\sqrt{2}b_2+b_3,\nonumber\\
 a_3&=&-3b_1+b_2-2(\sqrt{2}+i)b_3.
 \ea
Then, the canonical Hamiltonian can be reduced to the diagonal form
 \be
 {\cal H}_c=m\left(-b^\dg_1b_1+b^\dg_2b_2+b^\dg_3b_3\right).
 \ee
However, we are afraid to have unusual commutation relations between $b_i$ and
$b^\dg_j$
 \ba
 &&[b_1,b^\dg_1]=-\frac{133}{64},~~[b_1,b^\dg_2]=-\frac{55}{64},~~[b_1,b^\dg_3]=\frac{19\sqrt{2}}{32},\nonumber\\
 &&
 [b_2,b^\dg_2]=\frac{55}{64},~~~~~[b_2,b^\dg_3]=\frac{17\sqrt{2}}{32},~~[b_3,b^\dg_3]=\frac{3}{8},
 \ea
where all commutation relations survive. We call these unusual
commutators because all constant factors are nonstandard and signs
are negative in the first two expressions.

On the other hand, the canonical momenta for the ghost parts are
given by
 \be\label{canmomgh}
 \pi_c=\dot{d},~~~\pi_d=-\dot{c},
 \ee
and the canonical Hamiltonian is constructed by using modes
 \ba
  {\cal H}^{\rm gh}_c = m\left(d^\dg_1c_1-c^\dg_1d_1\right).
 \ea
Their anti-commutation relations are defined to be
 \be
 \{c,d^\dg\}=-1,~~\{d,c^\dg\}=1.
 \ee
Finally, the BRST charge can be obtained as
 \be
 Q_B=a^\dg_1c_1-c^\dg_1a_1,~~~~Q^2_B=0,
 \ee
and, the total canonical Hamiltonian of ${\cal H}_c+{\cal H}^{\rm
gh}_c$ is invariant under the BRST transformation of
 \ba
 &&\delta a_1=[Q_B,a_1]=0,~~\delta a_2=[Q_B,a_2]=0,~~\delta a_3=[Q_B,a_3]=c_1, \nonumber \\
 && \delta c_1=[Q_B,c_1]=0,~~\delta d_1=[Q_B,d_1]=a_1,
 \ea
which imply that $\{a_2\}$ represents unitary Fock space, while
$\{a_1,a_3,c_1,d_1\}$ form a quartet representation of BRST algebra.
Here we wish to point out that for dipole ghost fields (fourth-order
time derivative theory), the BRST invariant states appears only zero
norm combination through the quartet mechanism. This indicates  that
physical state is the vacuum~\cite{Rivelles:2003jd}. On the other
hand, for the sixth-order time derivative scalar theory, the
physical state is not the  vacuum but $\phi_2$. This shows clearly a
way of how a higher derivative harmonic oscillator is free from
negative norm states when quantizing it.
 Also, there is no imaginary
propagation speed that would signify a classical instability  in the
form of a tachyonic mode  appeared when replacing $m^2$ by $-m^2$ in
Eqs.~(\ref{eomp1})-(\ref{eomp3}).

At this stage, we are interested in the BRST quantization of free
theory (\ref{action2}) with ghosts.  The canonical and Dirac
quantizations of (\ref{action2}) without ghosts are beyond our scope
even though the presence of ghosts is harmful to the dynamics of
original system. For the canonical and Dirac quantizations of
Pais-Uhlenbeck oscillator, see
Refs.~\cite{Mannheim:2004qz,Smilga:2008pr}

\section{Quantization of sixth order derivative scalar theory }

In this section, we consider  the full action (\ref{action1}) in the
field theoretic point of view. Inspired by the BRST quantization in
the previous section, we wish to carry out the BRST quantization of
the field action (\ref{action1}). Usually, the BRST symmetry was
found in gauge theories as a symmetry of the gauge-fixed action. Its
purpose is definitely to remove the negative norm states associated
with the gauge invariance. Physical states are defined as those
which have zero ghost number and are invariant under the BRST
transformations. Here the BRST transformation (\ref{brstra}) is not
due to gauge invariance. Surely, it takes into account a feature of
giving  the sixth-order derivative structure starting from the
second-order action (\ref{action1}) via coupling. Since we have a
BRST symmetry, there is no doubt to require that the physical states
are those which have zero ghost number and are left invariant under
the BRST transformation.

Varying the fields $\phi_3$, $\phi_2$, and $\phi_1$ in the action
(\ref{action1}), we have equations of motion
 \ba
 \label{eomf1}\left(\nabla^2-m^2\right)\phi_1&=&0,\\
 \label{eomf2}\left(\nabla^2-m^2\right)\phi_2&=&\phi_1,\\
 \label{eomf3}\left(\nabla^2-m^2\right)\phi_3&=&\phi_2,
 \ea
respectively. Making use of an ansatz
 \be
 \phi_1(x)=\int\frac{d^3k}{(2\pi)^{3/2}\sqrt{2\omega}}\phi_{1k}(\vec{k},t)e^{i\vec{k}\cdot\vec{x}}
 \ee
the equation of motion (\ref{eomf1}) leads to one dimensional
equation for $ \phi_{1k}(\vec{k},t)$ as
 \be
 \left(\frac{d^2}{dt^2}+\omega^2\right)\phi_{1k}(\vec{k},t)=0,
 \ee
where $\omega^2=\vec{k}^2+m^2$. As like in the quantum mechanical
model in the previous section, this gives a solution expanded in
Fourier modes
 \be\label{phi1sol}
 \phi_1(x)=\int\frac{d^3k}{(2\pi)^{3/2}\sqrt{2\omega}} iN_1
       \left(a_1(\vec{k})e^{-i\omega t+i\vec{k}\cdot\vec{x}}-a^\dg_1(\vec{k})e^{i\omega t-i\vec{k}\cdot\vec{x}}\right).
 \ee
Similarly, we can find the solutions  to Eqs.~(\ref{eomf2})
and  (\ref{eomf3}) by using   (\ref{phi1sol}) as
 \begin{widetext}
 \ba\label{phi2sol}
 \phi_2(x)&=&\int\frac{d^3k}{(2\pi)^{3/2}\sqrt{2\omega}}\left(\frac{N_1}{2\omega^2}\right)
        \left[\left(a_2(\vec{k})+a_1(\vec{k})\omega t\right)e^{-i\omega t+i\vec{k}\cdot\vec{x}}
                    +\left(a^\dg_2(\vec{k})+a^\dg_1(\vec{k})\omega t\right)e^{i\omega
                    t-i\vec{k}\cdot\vec{x}}\right],\nonumber\\ \\
 \label{phi3sol}\phi_3(x)&=&\int\frac{d^3k}{(2\pi)^{3/2}\sqrt{2\omega}}\left(-\frac{iN_1}{4\omega^4}\right)
          \left[\left(a_3(\vec{k})+\left(a_2(\vec{k})-\frac{i}{2}a_1(\vec{k})\right)\omega t
                      +\frac{1}{2}a_1(\vec{k})\omega^2t^2\right)e^{-i\omega t+i\vec{k}\cdot\vec{x}}\right.\nonumber\\
     && ~~~~~~~~~~~~~~~~~~~~~~~ - \left.\left(a^\dg_3(\vec{k})+\left(a^\dg_2(\vec{k})+\frac{i}{2}a^\dg_1(\vec{k})\right)\omega t
                      +\frac{1}{2}a^\dg_1(\vec{k})\omega^2t^2\right)e^{i\omega t-i\vec{k}\cdot\vec{x}}\right].
 \ea
 \end{widetext}
On the other hand, for the ghost fields, their equations of motion
are given by
 \ba
 && \label{eomf4}\left(\nabla^2-m^2\right)c=0,\\
 && \left(\nabla^2-m^2\right)d=0,
 \ea
whose solutions are found to be
  \begin{widetext}
 \ba
 c(x)&=&\int\frac{d^3k}{(2\pi)^{3/2}\sqrt{2\omega}}\left(-\frac{iN_1}{4\omega^4}\right)
               \left(c_1(\vec{k})e^{-i\omega t+i\vec{k}\cdot\vec{x}}
                      -c^\dg_1(\vec{k})e^{i\omega  t-i\vec{k}\cdot\vec{x}}\right),\\
 d(x)&=&\int\frac{d^3k}{(2\pi)^{3/2}\sqrt{2\omega}}\left(iN_1\right)
               \left(d_1(\vec{k})e^{-i\omega t+i\vec{k}\cdot\vec{x}}
                      -d^\dg_1(\vec{k})e^{i\omega t-i\vec{k}\cdot\vec{x}}\right).
 \ea
 \end{widetext}
Now, canonical momenta is given by those as in Eq.~(\ref{canmom}),
and canonical Hamiltonian is obtained to be
 \ba
 {\cal H}_c = \int d^3x \left[\pi_1\pi_3+\frac{1}{2}\pi^2_2+\phi_1\phi_2+m^2\phi_1\phi_3+\frac{1}{2}m^2\phi^2_2
              +\vec{\nabla}\phi_1\cdot\vec{\nabla}\phi_3+\frac{1}{2}(\vec{\nabla}\phi_2)^2\right].
 \ea
It is  tedious  but straightforward to express the canonical
Hamiltonian in terms of  modes $a_1(\vec{k})$, $a_2(\vec{k})$, and
$a_3(\vec{k})$  as
 \ba
 {\cal H}_c&=&\int\frac{d^3k}{2\omega}\left(\frac{N^2_1}{2\omega^2}\right)
          \left[a^\dg_2(\vec{k})a_2(\vec{k})+2i\left(a^\dg_2(\vec{k})a_1(\vec{k})-a^\dg_1(\vec{k})a_2(\vec{k})\right)\right.\nonumber\\
          &&~~~~~~~~~~~~~~~~~~~~\left. -a^\dg_3(\vec{k})a_1(\vec{k}) -a^\dg_1(\vec{k})a_3(\vec{k}) \right],
 \ea
which corresponds to  the field theoretical  representation of the canonical
Hamiltonian (\ref{canH_ptls}).

Importantly, the commutation relations are given by
 \ba
 && [a_1(\vec{k}),a^\dg_3(\vec{k'})]=-2\omega\delta^3(\vec{k}-\vec{k'}),~~ [a_2(\vec{k}),a^\dg_2(\vec{k'})]=2\omega\delta^3(\vec{k}-\vec{k'}),\nonumber\\
 && [a_2(\vec{k}),a^\dg_3(\vec{k'})]=2i\omega\delta^3(\vec{k}-\vec{k'}),~~~\, [a_3(\vec{k}),a^\dg_3(\vec{k'})]=3\omega\delta^3(\vec{k}-\vec{k'}),
 \ea
where we have also taken $N^2_1=2\omega^3$. These can be also cast
into the  matrix form
 \be \label{scft}
 [a_i(\vec{k}), a^\dg_j(\vec{k'})]= 2\omega
 \left(
  \begin{array}{ccc}
   0 & 0 & -1 \\
    0 & 1 & i \\
    -1 & -i & \frac{3}{2} \\
  \end{array}
 \right)\delta^3(\vec{k}-\vec{k'}),
 \ee
which is similar to the two-point correlation functions in rank-3
LCFT~\cite{Bergshoeff:2012sc} \be \label{lcft}
 <{\cal O}^i{\cal O}^j> \sim
 \left(
  \begin{array}{ccc}
   0 & 0 & CFT \\
    0 & CFT & L \\
    CFT & L & L^2 \\
  \end{array}
 \right),
 \ee
where $i,j=$KG, log, log$^2$. Here  $CFT$ is the CFT correlation
function, $L$ represents log-correlation function, $L^2$ denotes
log$^2$-correlation function. A truncation to have a unitary
subspace is carried by throwing all modes which generate the third
column and row of this matrix containing $L^2$. Hence, the only
non-zero correlation function is proportional to the ordinary CFT
correlation. Using the AdS/LCFT correspondence, the remaining bulk
modes have a non-negative scalar product and the truncated theory is
unitary. This method can be generalized to arbitrary odd rank but it
fails for even rank  LCFTs.

By making use of the transformations of
 \ba
 a_1(\vec{k})&=&\frac{3\sqrt{2}i}{8}b_1(\vec{k})-\frac{\sqrt{2}i}{8}b_2(\vec{k})+\frac{i}{2}b_3(\vec{k}),\nonumber\\
 a_2(\vec{k})&=&\sqrt{2}b_1(\vec{k})-\sqrt{2}b_2(\vec{k})+b_3(\vec{k}),\nonumber\\
 a_3(\vec{k})&=&-3b_1(\vec{k})+b_2(\vec{k})-2(\sqrt{2}+i)b_3(\vec{k}),
 \ea
the canonical Hamiltonian can be successfully reduced to a
diagonal form
 \be
 {\cal H}_c=\int\frac{d^3k}{2\omega}
          \left[-b^\dg_1(\vec{k})b_1(\vec{k})+b^\dg_2(\vec{k})b_2(\vec{k})+b^\dg_3(\vec{k})b_3(\vec{k})\right].
 \ee
Canonical momenta for the ghosts are the same with
Eq.~(\ref{canmomgh}), and canonical Hamiltonian is given by
 \ba
 {\cal H}^{gh}_c =
 \int\frac{d^3k}{2\omega}~\omega\left(d^\dg_1(\vec{k})c_1(\vec{k})-c^\dg_1(\vec{k})d_1(\vec{k})\right).
 \ea
Their commutation relations are
 \ba
  \{c_1(\vec{k}),d^\dg_1(\vec{k'})\}=-2\omega\delta^3(\vec{k}-\vec{k'}),
  ~~\{d_1(\vec{k}),c^\dg_1(\vec{k'})\}=2\omega\delta^3(\vec{k}-\vec{k'}).
 \ea

Finally, the BRST charge can be obtained as the Noether charge
 \be
 Q_B=\int d^3x \left(a^\dg_1(\vec{k})c_1(\vec{k})-c^\dg_1(\vec{k})a_1(\vec{k})\right),~~~~Q^2_B=0,
 \ee
and the total canonical Hamiltonian of ${\cal H}_c+{\cal H}^{\rm
gh}_c$ is invariant under the BRST transformations
 \ba
 && \delta a_1(\vec{k})=[Q_B,a_1]=0,~~\delta a_2(\vec{k})=[Q_B,a_2]=0,
    ~~\delta a_3(\vec{k})=[Q_B,a_3]=c_1(\vec{k}),\nonumber\\
 && \delta c_1(\vec{k})=[Q_B,c_1]=0,~~\delta d_1(\vec{k})=[Q_B,d_1]=a_1(\vec{k}).
 \ea
Here we observe that $\{a_2\}$ is invariant under the BRST
transformations, while $\{a_1,a_3,c_1,d_1\}$ form a quartet
representation of BRST algebra to give zero norm state, which is
shown in the following:
 \be
 \label{quartet}
 [A_i,A^\dg_j]=2\omega\left(
  \begin{array}{c|ccccc}
     & ~a^\dg_2 & a^\dg_1 & a^\dg_3 & c^\dg &  d^\dg \\ \hline
   a_2 & 1    & 0  & i &  0 & 0 \\
   a_1 &~0   & 0  &-1 & 0  &  0 \\
   a_3 &~ i &-1  & 0 &  0 & 0 \\
   c & 0  & 0  & 0  &  0 & -1 \\
   d &  0 &  0 & 0  &  1 & 0 \\
  \end{array}
 \right)\delta^3(\vec{k}-\vec{k'}),
 \ee
where the $[\cdots]$ represents for commutator for
$a_i(a^\dagger_j)$, while it denotes anti-commutator for
$c,d(c^\dagger,d^\dagger)$. For the sixth-order derivative scalar
theory, the physical state is not vacuum but $\phi_2$. This shows
clearly a way of how a higher derivative scalar field theory is free
from negative norm states. Comparing it with Yang-Mills theory
(4.52) in~\cite{Kugo:1979gm}, we have apparent correspondence
between two
\begin{equation}
a_1 \leftrightarrow B,~~a_2 \leftrightarrow A_T,~~a_3
\leftrightarrow A_L,
\end{equation}
where $B$ is a conjugate momentum of scalar gauge mode $A_S$, while
$A_T$ represents the transverse gauge mode with positive norm and
$A_L$ denotes the longitudinal gauge mode. Also, a difference comes
from a non-zero commutator of $[a_2,a^\dagger_3]=2i\omega
\delta^3(\vec{k}-\vec{k'})$ which plays an important role in
truncation procedure when comparing it with  rank-3 LCFT
(\ref{lcft}).

\section{Discussions}

We have clarified  the truncation scheme to provide the unitary
subspace in the rank-3 LCFT. This was  an obtaining procedure of a
unitary subspace after forming a BRST quartet even for a sixth-order
derivative  scalar field theory.  The correspondence between
 the rank-3 LCFT and sixth order derivative scalar field theory is given
by observing (\ref{scft}) and (\ref{lcft}). This is clear because
(\ref{lcft}) was obtained  from the same action (\ref{action1}) on
the boundary  of the AdS$_3$ spacetime without ghosts. The
difference is that we consider (\ref{action1}) in Minkowski
spacetime.

 Finally, in this work, we did not fully perform the canonical
and Dirac quantizations of (\ref{action2}) without ghosts which
seems to be a formidable task.

\begin{acknowledgments}
This work was supported by the National Research Foundation of Korea
(NRF) grant funded by the Korea government (MISP) through the Center
for Quantum Spacetime (CQUeST) of Sogang University with grant
number 2005-0049409. Y. S. Myung was also supported by the National
Research Foundation of Korea (NRF) grant funded by the Korea
government (MISP) (No.2012-R1A1A2A10040499).
\end{acknowledgments}



\begin{thebibliography}{99}

\bibitem{Stelle}
  K.~S.~Stelle,
  Phys.\ Rev.\  D {\bf 16}, 953 (1977).

\bibitem{Barth:1983hb}
  N.~H.~Barth and S.~M.~Christensen,
  Phys.\ Rev.\ D {\bf 28}, 1876 (1983).



\bibitem{Li:2008dq}
  W.~Li, W.~Song and A.~Strominger,
   JHEP {\bf 0804}, 082 (2008)  [arXiv:0801.4566 [hep-th]].

\bibitem{Lu:2011zk}
  H.~Lu and C.~N.~Pope,
  Phys.\ Rev.\ Lett.\  {\bf 106}, 181302 (2011)  [arXiv:1101.1971 [hep-th]].

\bibitem{Deser:2011xc}
  S.~Deser, H.~Liu, H.~Lu, C.~N.~Pope, T.~C.~Sisman and B.~Tekin,
  Phys.\ Rev.\ D {\bf 83}, 061502 (2011)  [arXiv:1101.4009 [hep-th]].


\bibitem{Alishahiha:2011yb}
  M.~Alishahiha and R.~Fareghbal,
  Phys.\ Rev.\ D {\bf 83}, 084052 (2011)  [arXiv:1101.5891 [hep-th]].

\bibitem{Bergshoeff:2011ri}
  E.~A.~Bergshoeff, O.~Hohm, J.~Rosseel and P.~K.~Townsend,
  Phys.\ Rev.\ D {\bf 83}, 104038 (2011)  [arXiv:1102.4091 [hep-th]].

\bibitem{Porrati:2011ku}
  M.~Porrati and M.~M.~Roberts,
   Phys.\ Rev.\ D {\bf 84}, 024013 (2011)  [arXiv:1104.0674 [hep-th]].


\bibitem{Myung:2011bn}
  Y.~S.~Myung, Y.~-W.~Kim, T.~Moon and Y.~-J.~Park,
  Phys.\ Rev.\ D {\bf 84}, 024044 (2011)  [arXiv:1105.4205 [hep-th]].

\bibitem{Lu:2011mw}
  H.~Lu, C.~N.~Pope, E.~Sezgin and L.~Wulff,
  JHEP {\bf 1110}, 131 (2011)  [arXiv:1107.2480 [hep-th]].


\bibitem{Grumiller:2008qz}
  D.~Grumiller and N.~Johansson,
   JHEP {\bf 0807}, 134 (2008)  [arXiv:0805.2610 [hep-th]].
\bibitem{Myung:2008dm}
  Y.~S.~Myung,
  Phys.\ Lett.\ B {\bf 670}, 220 (2008)  [arXiv:0808.1942 [hep-th]].

\bibitem{Maloney:2009ck}
  A.~Maloney, W.~Song and A.~Strominger,
  Phys.\ Rev.\ D {\bf 81}, 064007 (2010)  [arXiv:0903.4573 [hep-th]].

\bibitem{Gurarie:1993xq}
  V.~Gurarie,
    Nucl.\ Phys.\ B {\bf 410}, 535 (1993)  [hep-th/9303160].

\bibitem{Flohr:2001zs}
  M.~Flohr,
   Int.\ J.\ Mod.\ Phys.\ A {\bf 18}, 4497 (2003)  [hep-th/0111228].

\bibitem{Bergshoeff:2012sc}
  E.~A.~Bergshoeff, S.~de Haan, W.~Merbis, M.~Porrati and J.~Rosseel,
   JHEP {\bf 1204}, 134 (2012)  [arXiv:1201.0449 [hep-th]].


\bibitem{Kogan:1999bn}
  I.~I.~Kogan,
   Phys.\ Lett.\ B {\bf 458}, 66 (1999)  [hep-th/9903162].


\bibitem{Myung:1999nd}
  Y.~S.~Myung and H.~W.~Lee,
  JHEP {\bf 9910}, 009 (1999)  [hep-th/9904056].


\bibitem{Becchi:1975nq}
  C.~Becchi, A.~Rouet and R.~Stora,
  Annals Phys.\  {\bf 98}, 287 (1976).

\bibitem{Tyutin:1975qk}
  I.~V.~Tyutin,
  LEBEDEV-75-39,
  arXiv:0812.0580 [hep-th].


\bibitem{Kugo:1979gm}
  T.~Kugo and I.~Ojima,
   Prog.\ Theor.\ Phys.\ Suppl.\  {\bf 66}, 1 (1979).

\bibitem{Rivelles:2003jd}
  V.~O.~Rivelles,
  Phys.\ Lett.\ B {\bf 577}, 137 (2003)  [hep-th/0304073].



\bibitem{Nutma:2012ss}
  T.~Nutma,
  Phys.\ Rev.\ D {\bf 85}, 124040 (2012)  [arXiv:1203.5338 [hep-th]].


\bibitem{Kleinschmidt:2012rs}
  A.~Kleinschmidt, T.~Nutma and A.~Virmani,
  Gen.\  Rel.\  Grav.\  {\bf 45}, 727 (2013)  [arXiv:1206.7095 [hep-th]].



\bibitem{Bergshoeff:2012ev}
  E.~A.~Bergshoeff, S.~de Haan, W.~Merbis, J.~Rosseel and T.~Zojer,
   Phys.\ Rev.\ D {\bf 86}, 064037 (2012)  [arXiv:1206.3089 [hep-th]].



\bibitem{Moon:2012vc}
  T.~Moon and Y.~S.~Myung,
  Phys.\ Rev.\ D {\bf 86}, 084058 (2012)  [arXiv:1208.5082 [hep-th]].

\bibitem{de Urries:1998bi}
  F.~J.~de Urries and J.~Julve,
  J.\ Phys.\ A {\bf 31}, 6949 (1998)  [hep-th/9802115].

\bibitem{Pais:1950za}
  A.~Pais and G.~E.~Uhlenbeck,
   Phys.\ Rev.\  {\bf 79}, 145 (1950).


\bibitem{Jimenez:2012ak}
  J.~B.~Jimenez, E.~Dio and R.~Durrer,
  JHEP {\bf 1304}, 030 (2013)  [arXiv:1211.0441 [hep-th]].

\bibitem{Mannheim:2004qz}
  P.~D.~Mannheim and A.~Davidson,
  Phys.\ Rev.\ A {\bf 71}, 042110 (2005)  [hep-th/0408104].

\bibitem{Smilga:2008pr}
  A.~V.~Smilga,
   SIGMA {\bf 5}, 017 (2009)  [arXiv:0808.0139 [quant-ph]].


 \end{thebibliography}
 \end{document}